# Realization of a Laughlin quasiparticle interferometer: Observation of fractional statistics

F. E. Camino, Wei Zhou and V. J. Goldman

*Department of Physics, Stony Brook University, Stony Brook, New York 11794-3800, USA*

In two dimensions, the laws of physics permit existence of anyons, particles with fractional statistics which is neither Fermi nor Bose. That is, upon exchange of two such particles, the quantum state of a system acquires a phase which is neither 0 nor $\pi$, but can be *any* value. The elementary excitations (Laughlin quasiparticles) of a fractional quantum Hall fluid have fractional electric charge and are expected to obey fractional statistics. In this paper we report experimental realization of a novel Laughlin quasiparticle interferometer, where quasiparticles of the 1/3 fluid execute a closed path around an island of the 2/5 fluid and thus acquire statistical phase. Interference fringes are observed as conductance oscillations as a function of magnetic flux, similar to the Aharonov-Bohm effect. We observe the interference shift by one fringe upon introduction of five magnetic flux quanta ($5h/e$) into the island. The corresponding $2e$ charge period is confirmed directly in calibrated gate experiments. These results constitute direct observation of fractional statistics of Laughlin quasiparticles.

# I. Introduction

It has been long understood theoretically that in two spatial dimensions the laws of physics do not prohibit existence of particles with fractional exchange statistics, dubbed *anyons*.[1,2] This is because in two dimensions (2D) a closed loop executed by a particle around another particle is topologically distinct from a loop which encloses no particles, unlike the three dimensional case. The particles are said to have statistics $\Theta$ if upon exchange the two-particle wave function acquires a phase factor of $\exp(i\pi\Theta)$, and, upon a closed loop, a factor of $\exp(i2\pi\Theta)$. An exchange of two particles is equivalent to one particle executing a half loop around the other, so that a closed loop is equivalent to exchange squared. The integer values $\Theta_B = 2j$ and $\Theta_F = 2j+1$, where $j = 0, \pm 1, \pm 2, \ldots$, describe the familiar boson and fermion exchange statistics: $\exp(i2\pi j) = (-1)^{2j} = +1$ and $\exp[i\pi(2j+1)] = (-1)^{2j+1} = -1$, respectively. Upon execution of a closed loop both bosons and fermions produce a phase factor of $+1$, which is unobservable, so usually the statistical contribution can be safely neglected when describing an interference experiment, such as the Aharonov-Bohm effect.[3]

The fundamental "elementary" particles exist in three spatial dimensions, and thus all have either bosonic or fermionic integer statistics. Any particles having a fractional statistics must be elementary collective excitations of a nontrivial system of many integer statistics particles confined to move in 2D. Thinking in terms of a few of such weakly-interacting, fractional effective particles instead of in terms of very complex collective motions of all the underlying strongly-interacting, integer statistics particles greatly simplifies description of relevant physics. In particular, the elementary charged excitations (Laughlin quasiparticles[4]) of a fractional quantum Hall (FQH) electron fluid[5,4] have a fractional electric charge[4,6] and therefore are expected to obey fractional statistics.[7,8]

Arovas, Schrieffer and Wilczek[8] have used the adiabatic theorem to calculate the Berry phase[9] $\gamma$ of a charge $e/3$ Laughlin quasiparticle at position $\Re$ encircling a closed path $C$ containing another $e/3$ quasiparticle at $\Re'$ in the filling $f = 1/3$ FQH condensate:

$$\gamma = i \oint_C d\Re \left\langle \Psi(\Re, \Re') \left| \frac{\partial}{\partial \Re} \Psi(\Re, \Re') \right\rangle \right., \tag{1}$$

where $\Psi$ is the many-electron Laughlin wave function.[4] They found the difference,

$$\Delta\gamma = 2\pi \Theta_{1/3} = 4\pi/3, \tag{2}$$

identified as the statistical contribution, between an "empty" loop and a loop containing another quasiparticle. It is possible to assign definite fractional statistics (mod 1) to quasiparticles of certain simple FQH fluids based only on the knowledge of their charge.[10] For example, for the one electron layer FQH fluids corresponding to the main composite fermion sequence[11,12] $f = p/(2jp+1)$, with $p$ and $j$ positive integers, the charge $q = e/(2jp+1)$ quasiparticle statistics is expected to be

$$\Theta_{p/(2jp+1)} = 2j/(2jp+1) \text{ (mod 1)}. \tag{3}$$

It is instructive to consider a simple example. A quantum antidot electrometer has been used in the direct observation of the charge $e/3$ and $e/5$ quasiparticles,[6,13,14] subsequently confirmed in shot noise measurements.[15,16] A quantum antidot is a potential hill lithographically defined in a 2D electron layer in the quantum Hall regime. The wave functions of a charge $q$ particle encircling the antidot are quantized by the Aharonov-Bohm condition (explicitly including the statistical contribution):

$$\gamma_m = \frac{q}{\hbar} \Phi + 2\pi \Theta N = 2\pi m, \tag{4}$$

where $m$ is an integer, $\Phi$ is the enclosed flux and $N$ is the number of antidot-bound quasiholes being encircled.[17] When the chemical potential $\mu$ moves between two successive quasiparticle states, the change in the phase of the wave function is $2\pi$:

$$\Delta\gamma \equiv \gamma_{m+1} - \gamma_m = \frac{q}{\hbar}\Delta\Phi + 2\pi\Theta\,\Delta N = \pm 2\pi\,. \qquad (5)$$

When occupation of the antidot changes by one $e/3$ quasiparticle, $\Delta N = 1$, the experiments[6,14] give $\Delta\Phi = h/e$, so that $\Delta\gamma = 2\pi(q/e + \Theta_{1/3}) = 2\pi$ only if quasiparticles have a fractional $\Theta_{1/3} = 2/3$.

This experiment, however, is not entirely satisfactory as a *direct* demonstration of the fractional statistics of Laughlin quasiparticles because in a quantum antidot the tunneling quasiparticle encircles electron vacuum. Thus, the most important ingredient, the experimental fact that in quantum antidots the period $\Delta\Phi = h/e$, and not $\Delta\Phi = h/q$, even for fractionally charged particles, is ensured by the gauge invariance argument of the Byers-Yang theorem.[18] Several theoretical studies pointed out that fractional statistics of Laughlin quasiparticles can be observed experimentally in variations of the Aharonov-Bohm effect,[19-21] but the experimental evidence has been lacking.

Our present experiment utilizes a novel Laughlin quasiparticle interferometer, where a quasiparticle with charge $e/3$ of the $f = 1/3$ FQH fluid executes a closed path around an island of the $f = 2/5$ fluid, see Figure 1. The interference fringes are observed as peaks in conductance as a function of the magnetic flux through the $f = 2/5$ island, in a kind of the Aharonov-Bohm effect. We observe the Aharonov-Bohm period of five magnetic flux quanta through the $f = 2/5$ island, i.e. $\Delta\Phi = 5h/e$, corresponding to excitation of ten $q = e/5$ quasiparticles of the $f = 2/5$ fluid. Such "superperiod" of $\Delta\Phi > h/e$ has never been reported before. The corresponding $\Delta Q = 10(e/5) = 2e$ charge period is directly confirmed in calibrated backgate experiments. These observations imply *relative* statistics of $\Theta^{1/3}_{2/5} = -1/15$, when a charge $e/3$ Laughlin quasiparticle encircles one $e/5$ quasiparticle of the $f = 2/5$ fluid.

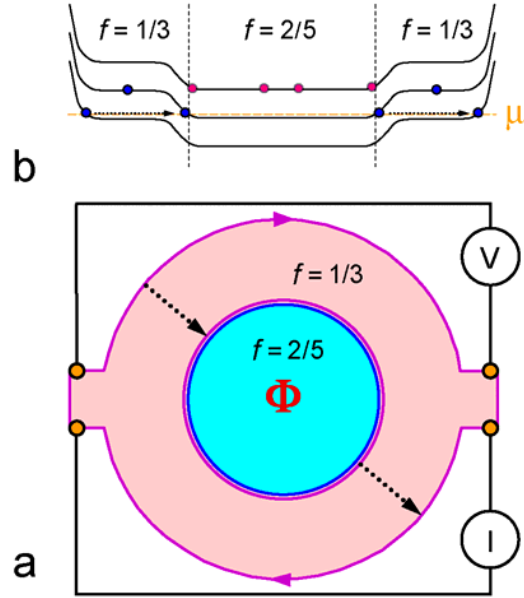

**Fig. 1** (Color online) Conceptual schematic of the Laughlin quasiparticle interferometer. (a) A quantum Hall sample with two fillings: an island of 2/5 FQH fluid is surrounded by the 1/3 fluid. The current-carrying chiral edge channels (shown by arrowed magenta lines) follow equipotentials at the periphery of the confined 2D electrons; tunneling paths are shown by dots. The orange circles are the Ohmic contacts used to inject current $I$ and to measure the resulting voltage $V$. The central island is encircled by two counterpropagating edge channels. The current-carrying $e/3$ quasiparticles can tunnel between the outer and the inner 1/3 edges, dotted lines. When there is no tunneling, $V = 0$; tunneling produces $V > 0$. The closed path of the *inner* 1/3 edge channel gives rise to Aharonov-Bohm-like oscillations in conductance as a function of the enclosed flux $\Phi$. No current flows through the 2/5 island, but any $e/5$ quasiparticles affect the Berry phase of the encircling $e/3$ quasiparticles through a statistical interaction, thus changing the interference pattern. (b) FQH liquids can be understood via composite fermion representation. A composite fermion energy profile of the interferometer shows the three lowest "Landau levels" separated by FQH energy gaps. Several $e/3$ and $e/5$ quasiparticles are shown as composite fermions in otherwise empty "Landau levels".

## II. Experimental technique

The quantum electron interferometer samples were fabricated from a low disorder GaAs/AlGaAs heterojunction material where 2D electrons (285 nm below the surface) are prepared by exposure to red light at 4.2 K. The four independently-contacted front gates were defined by electron beam lithography on a pre-etched mesa with Ohmic contacts. After a shallow 140 nm wet chemical etching, Au/Ti gate metal was deposited in etch trenches, followed by lift-off, see Figure 2(a, b). Samples were mounted on sapphire substrates with In metal, which serves as a global backgate. Samples were cooled to 10.2 mK in the mixing chamber tail of a top-loading into mixture $^3$He-$^4$He dilution refrigerator. Four-terminal resistance $R_{XX} \equiv V_X/I_X$ was measured by passing a 100 pA, 5.4 Hz *ac* current through contacts 1 and 4, and detecting the voltage between contacts 2 and 3 by a lock-in-phase technique. An extensive cold filtering cuts the integrated electromagnetic "noise" environment incident on the sample to $\sim 5 \times 10^{-17}$ W, which allows us to achieve a record low electron temperature of 18 mK in a mesoscopic sample.[22]

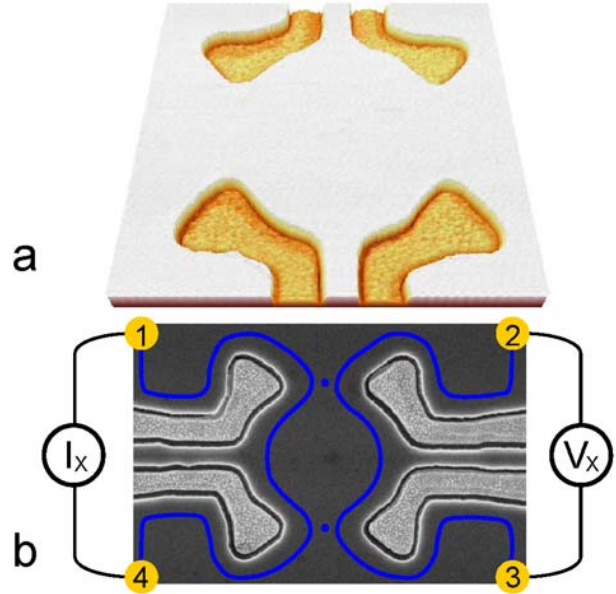

**Fig. 2** (Color online) The Laughlin quasiparticle interferometer samples. (a) and (b) are atomic force and scanning electron micrographs of a typical device. Four Au/Ti front gates in shallow etch trenches define the central island of 2D electrons of lithographic radius $R \approx 1,050$ nm. The 2D electrons are completely depleted under the etched trenches; the front gates are used for fine-tuning the two wide constrictions. The chiral edge channels (blue) follow equipotentials at the periphery of the undepleted 2D electrons; tunneling paths are shown by dots. Four Ohmic contacts are shown schematically by the numbered circles, $R_{XX} \equiv V_{2-3}/I_{1-4}$. The backgate (not shown) extends over the entire sample on the opposite side of the insulating GaAs substrate.

The four front gates are deposited into etch trenches. In this work, the voltages applied to the four front gates $V_{FG}$ (with respect to the 2D electron layer) are small, and are used to fine tune for the symmetry of the two constrictions. Even when front gate voltages $V_{FG} = 0$, the GaAs surface depletion potential of the etch trenches defines two wide constrictions, which separate an approximately circular 2D electron island with lithographic radius $R \approx 1{,}050$ nm from the 2D "bulk". The electron density profile $n(r)$ in a circular island resulting from the etch trench depletion can be evaluated following the model of Gelfand & Halperin,[23] see Figure 3. For the 2D bulk electron density $n_B = 1.2 \times 10^{11}$ cm$^{-2}$, there are ~2,000 electrons in the island. Under such conditions ($V_{FG} \approx 0$), the depletion potential has a saddle point in the constriction region, and so has the resulting electron density profile. From the magnetotransport measurements (see below) we estimate the saddle point density value $n_C \approx 0.75 n_B$, which varies somewhat due to the self-consistent electrostatics of the 2D electrons in presence of a quantizing magnetic field. Note that the island center density is slightly (several per cent) lower than the 2D bulk density.

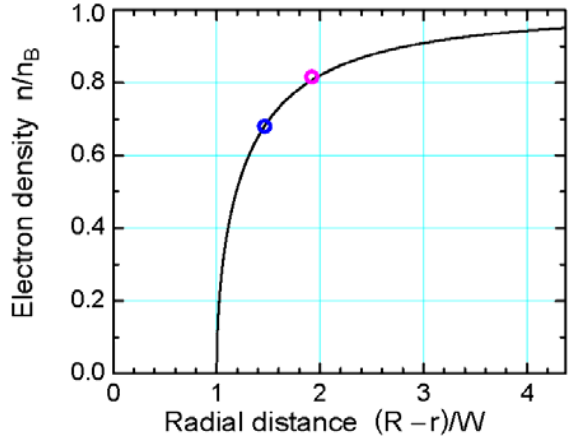

**Fig. 3** (Color online) The electron density $n$, normalized to the 2D density, as a function of radius $r$ in an electron island defined by an etched annulus of inner radius $R \approx 1{,}050$ nm. The calculation follows model of Ref. 23. $W = 250$ nm is the depletion length parameter obtained in the same calculation. The blue circle gives the radius of the outer edge ring, $r_{Out} \approx 685$ nm, obtained from the Aharonov-Bohm period data for $f_B = 1$ and 2, shown in Figure 6. The magenta circle gives the radius of the inner edge ring, $r_{In} \approx 570$ nm for $f_B = 2/5$ from the data of Figs. 9 and 10, taking the same $r_{Out}$ for $f_C = 1/3$, and the ratio of densities $n(r_{In})/n(r_{Out}) = (2/5)/(1/3) = 1.20$.

**III. Integer QH regime**

Here the relevant quasiparticles are electrons of charge $e$ and integer statistics, therefore, we can obtain an absolute calibration of the ring area and the backgate action of the interferometer device. Figure 4 shows the directly measured four-terminal $R_{XX}$ as a function of applied normal magnetic field $B$. The *local* Landau level filling factor $\nu \equiv hn/eB$ is proportional to $n(r)$, and the electron density in the constrictions $n_C < n_B$. Consequently the constriction $\nu_C$ is lower than the bulk $\nu_B$ by some 20 to 30% in a given $B$. While $\nu \propto n(r)/B$ is a variable, the quantum Hall exact filling $f$, defined as the inverse of value of the *quantized* Hall resistance $R_{XY}$ in units of $h/e^2$ (that is, $f \equiv h/e^2 R_{XY}$), is a quantum number. The 2D electron system on a quantum Hall plateau $f$ opens an energy gap. Variation of $B$ from the exact $\nu = f$ is accommodated by creation of quasiparticles ($\nu > f$) or quasiholes ($\nu < f$). Note that the 2D electronic charge $-en$ is only redistributed: $e(f - \nu)B/h$ is the charge density of the quasiparticles and $-efB/h$ of the condensate, the total system (2D electrons and donors) remains neutral. Analogous considerations apply when a global back gate is biased at a fixed $B$.

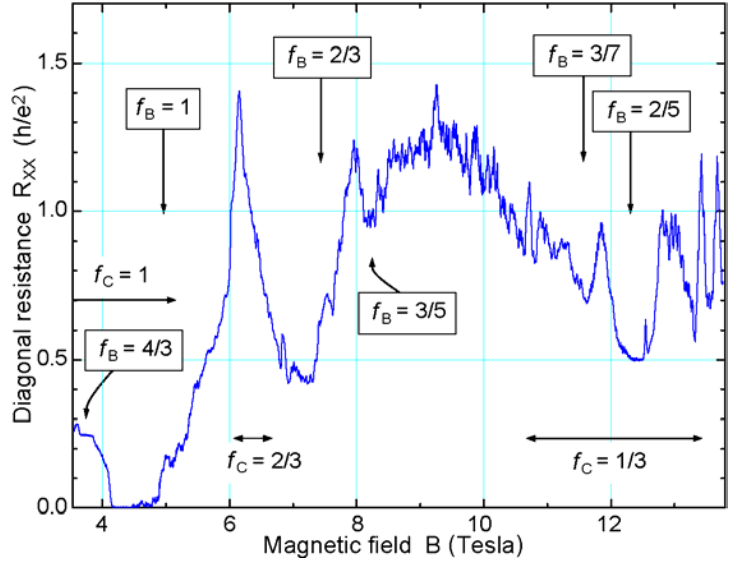

**Fig. 4** (Color online) Magnetoresistance of the quasiparticle interferometer sample at temperature 10.2 mK. The horizontal arrows show approximate ranges of filling factor $\nu_C = f_C$ plateaus in the constrictions. Note the *quantized* plateaus $R_{XX}(B) = h/4e^2$ at ~3.7 Tesla ($f_C = 1$, $f_B = 4/3$) and $R_{XX}(B) = h/2e^2$ at ~12.4 Tesla ($f_C = 1/3$, $f_B = 2/5$). The overlap of the $f_C = 1$ and $f_B = 1$ plateaus for 4.2 T $< B <$ 4.8 T likewise results in $R_{XX}(B) = 0$ in this range. The data were obtained with front gate voltage $V_{FG} = 0$, the front gates were not tuned for symmetry. The fine structure is due to quantum interference effects; some peaks can be identified as due to impurity-assisted tunneling.

Thus there are two quantum Hall regimes possible: one when the whole sample has one and the same quantum Hall filling $f$, and another when there are two quantum Hall fillings: $f_C$ in the constrictions, and $f_B$ in the center of the island and in the 2D bulk. For example, there is a range of $B$ such that both $f_C = 1$ and $f_B = 1$, as seen for 4.2 T $< B <$ 4.8 T in Figure 4, illustrated schematically in Figure 5(a). The second possibility is illustrated in Figures 5(b,c). For example, $f_C = 1$ and $f_B = 4/3$, resulting in a *quantized* value[24,25,6,14] of $R_{XX} = (h/e^2)(1/f_C - 1/f_B)$, is seen in the range 3.75 T $< B <$ 3.85 T in Figure 4. However, $f_C = 1$, $f_B = 2$, which would require $n_C \approx (1/2) n_B$, is not possible in this sample. Thus an observation of a quantized plateau in $R_{XX}(B)$ implies quantum Hall plateaus for both the constriction region and the bulk, and in practice provides definitive values for both $f_C$ and $f_B$, since the number of well-defined quantum Hall states $f_B$ observed in a given sample is usually rather finite.

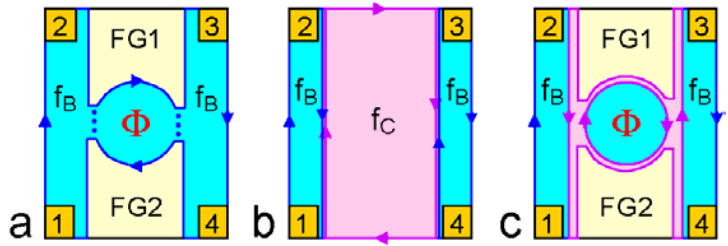

**Fig. 5** (Color online) (a) Schematic of the sample when magnetic field is such that there is only one quantum Hall filling $f$ throughout the sample: $f_C$ in the constrictions is equal to $f_B$ in the 2D bulk and in the island. The numbered rectangles are Ohmic contacts, FG are the front gates. The chiral edge channels follow equipotentials at the periphery of the undepleted 2D electrons; tunneling paths are shown by dots. A closed edge channel path gives rise to Aharonov-Bohm oscillations in the conductance. (b) A sample with two quantum Hall fillings exhibits quantized diagonal resistance $R_{XX} = (h/e^2)(1/f_C - 1/f_B)$, where $R_{XX} \equiv V_{2-3}/I_{1-4}$. Observation of a quantized plateau in $R_{XX}(B)$ provides definitive values for both $f_C$ and $f_B$. (c) Schematic of the sample when magnetic field is such that $f_C < f_B$. The sample exhibits a quantized $R_{XX}(B)$ plateau, and, upon fine tuning of front gates, exhibits Aharonov-Bohm oscillations in conductance as a function of the flux enclosed by the *inner* island edge ring.

In the integer quantum Hall regime we observe Aharonov-Bohm type conductance oscillations for $f_C = f_B$ = 1 and 2, see Figure 6. The oscillatory conductance variation $\delta G = \delta R_{XX}/R_{XY}^2$ is obtained[6,14] from the directly measured $R_{XX}$ data after subtracting a smooth background. The Aharonov-Bohm ring is formed here by the edge channel circling the island, and includes two quantum tunneling links, see Figure 5(a). The $f = 1$ Aharonov-Bohm period $\Delta B_1 \approx 2.81$ mT gives the area of the island "outer" edge ring $S_{Out} = h/e\Delta B_1 \approx 1.47$ μm², and the outer ring radius $r_{Out} = \sqrt{h/\pi e \Delta B_1} \approx 685$ nm. The $f = 2$ period is very close: $2\Delta B_2 \approx 2.85$ mT, and gives the area $S_{Out} \approx 1.45$ μm² and the radius $r_{Out} \approx 680$ nm. The $f = 2$ Aharonov-Bohm period contains two conductance oscillations, $\Delta\Phi_2 = 2\Delta B_2 S_O$, because there are two filled (spin-polarized) Landau levels, as reported previously for a constricted Coulomb island[26] and for a quantum antidot in the integer quantum Hall regime.[27]

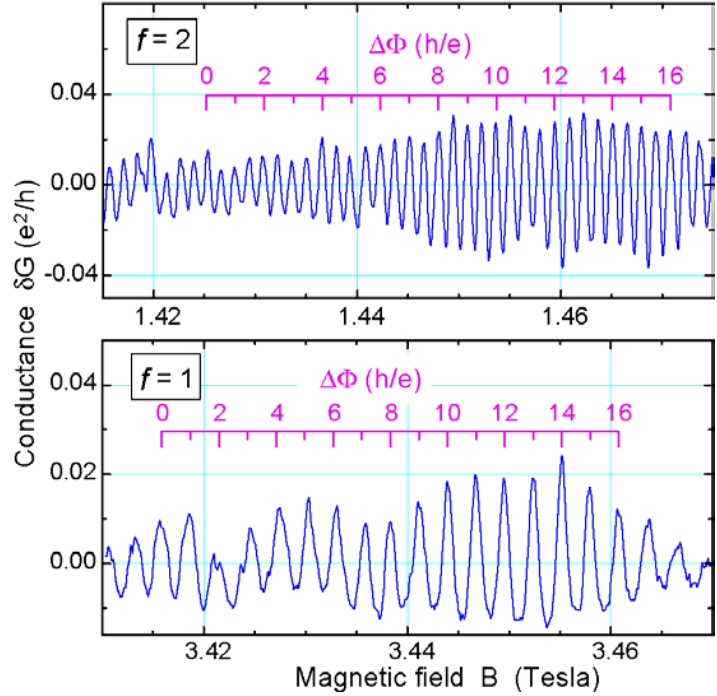

**Fig. 6** (Color online) Interference of electrons in the outer ring of the device in the integer quantum Hall regime. Aharonov-Bohm type oscillations in conductance are observed when one ($f=1$) and two ($f=2$) Landau levels are filled. The corresponding flux period $\Delta\Phi = h/e$ gives the outer ring radius $r_{Out} \approx 685$ nm.

In the integer quantum Hall regime, where elementary excitations are electrons in partially-filled Landau levels, we calibrate the global backgate. Applying positive $V_{BG}$ attracts electrons to the 2D layer, $V_{BG} = 1$ V increases $n_B$ by $\delta n_B \approx 2.4 \times 10^8$ cm$^{-2}$. The ratio $\delta n_B / n_B \approx 2.0 \times 10^{-3}$ is small since the backgate is separated from the 2D layer by a rather thick insulating GaAs substrate. Unlike the case of a quantum antidot,[6,14] where the antidot is completely surrounded by a quantum Hall condensate, we do not expect the density in the interferometer island to increase by exactly the same amount as in the 2D bulk. Therefore we must calibrate the $dQ/dV_{BG}$ ratio, where $Q$ is the charge of electrons within the island edge ring. Figure 7 shows the conductance oscillations observed as a function of $V_{BG}$ in the integer QH regime, $f_C = f_B = 1$ and 2. The period of these oscillations $\Delta V_{BG}$ corresponds to change of the number of electrons $N$ within the island edge ring by one. $\Delta V_{BG}$ is expected to be the same for all spin-polarized integer quantum Hall states, provided the radius of the edge ring does not change; indeed, we obtain $\Delta V_{BG} = 332$ mV for $f = 1$ and 342 mV for $f = 2$.

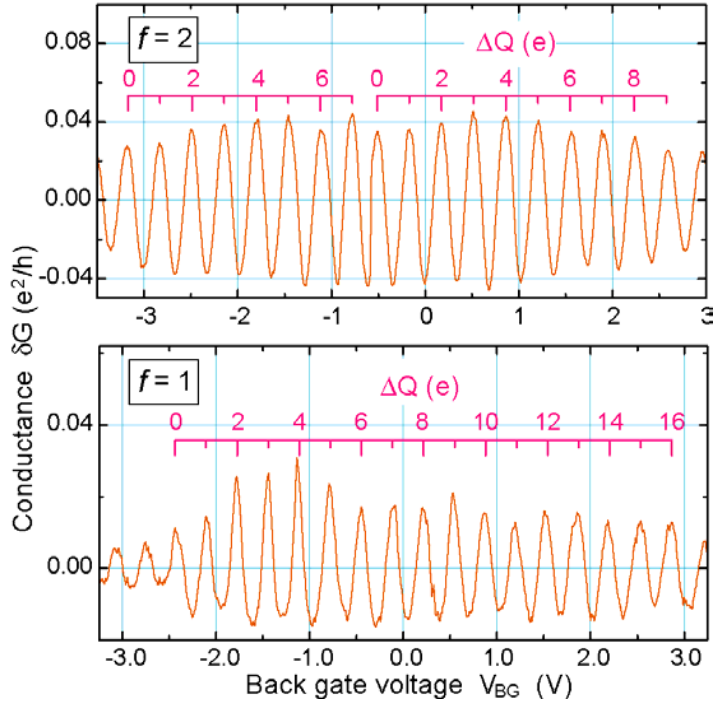

**Fig. 7** (Color online) Interference of electrons in the outer ring of the device in the integer quantum Hall regime. Application of a positive back gate voltage $V_{BG}$ attracts the 2D electrons one by one to the area of the outer ring, resulting in modulation of the interference amplitude. This calibrates the backgate voltage increment $\Delta V_{BG}$ necessary to increase the charge contained within the ring by $\Delta Q = e$. Note that $\Delta V_{BG}$ is independent of Landau level filling.

## IV. Fractional QH regime

Setting the applied $B$ such that the 2D bulk is on the $f_B = 2/5$ FQH plateau, we focus on the situation when an $f_C = 1/3$ annulus surrounds an island of the $f_B = 2/5$ FQH fluid, shown schematically in Figures 1 and 5(c). We can be confident that an $f_C = 1/3$ region separates the two 2D bulk $f_B = 2/5$ regions with Ohmic contacts because the diagonal resistance is quantized to $R_{XX} = (h/e^2)(3/1 - 5/2) = (1/2)(h/e^2)$, see Figure 8. Note that the $R_{XX}(B)$ gross structure (the peaks between the FQH plateaus) comes from the 2D bulk, not from the island. The island center electron density is slightly (several percent, see Fig. 3) less than the 2D bulk density, thus, in a given applied magnetic field, the island center filling factor is several percent lower than the 2D filling factor.

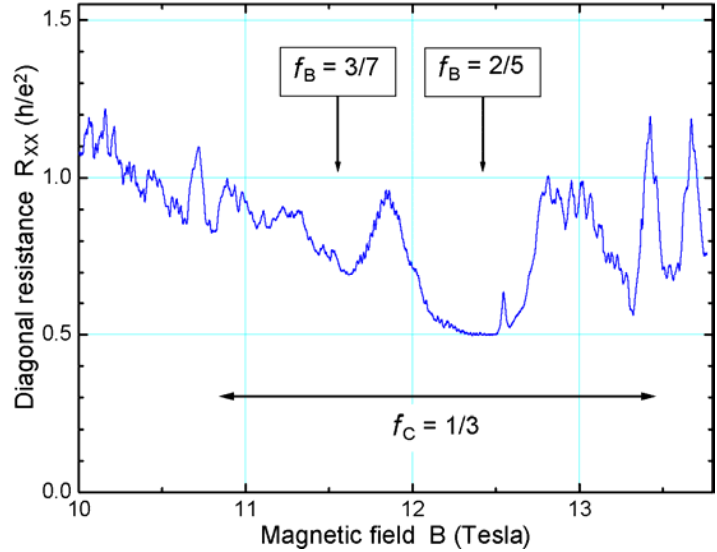

**Fig. 8** (Color online) A blow-up of the data of Figure 4. The horizontal arrows show approximate range of the $f_C = 1/3$ plateau in the constrictions. Note the *quantized* plateau $R_{XX}(B) = (h/e^2)[(1/3)^{-1} - (2/5)^{-1}] = h/2e^2$ at ~12.35 Tesla ($f_C = 1/3$, $f_B = 2/5$). The data were obtained with $V_{FG} = 0$, the front gates were not tuned for symmetry. The fine structure is due to quantum interference effects; some peaks can be identified as due to impurity-assisted tunneling.

Here, as in the integer regime, we also observe Aharonov-Bohm type conductance oscillations as a function of $B$, with period $\Delta B \approx 20.1$ mT, see Figs. 9 and 10(a). The corresponding flux period is $\Delta \Phi = 5h/e$. The oscillation period in this regime gives the inner edge ring area $S_{In} = 5h/e\Delta B \approx 1.03$ μm², and the inner ring radius $r_{In} = \sqrt{5h/\pi e \Delta B} \approx 570$ nm.

Figure 10(b) shows the analogous conductance oscillations observed as a function of $V_{BG}$, with period $\Delta V_{BG} = 937$ mV. The corresponding charge period is $\Delta Q = 2e$.

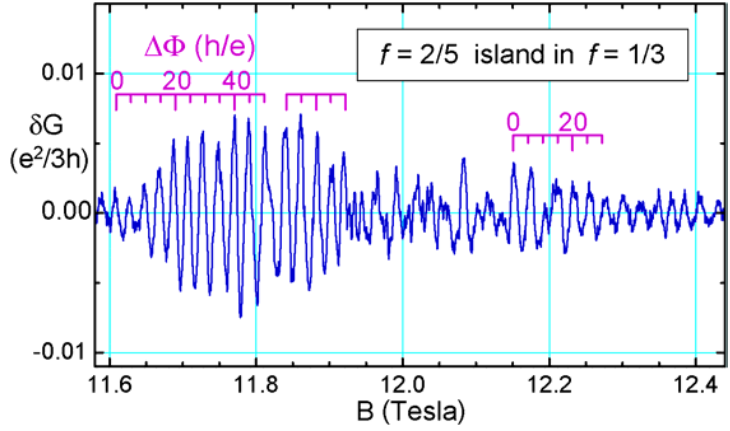

**Fig. 9** (Color online) Aharonov-Bohm effect of the inner ring $e/3$ Laughlin quasiparticles circling an island of the $f = 2/5$ FQH fluid. Magnetic flux through the island period of $\Delta \Phi = 5h/e$ corresponds to creation of ten $e/5$ quasiparticles in the island (one fundamental flux quantum $h/e$ induces two quasiparticles in the $f = 2/5$ FQH fluid). The red arrow gives the magnetic field where the backgate data were taken, Fig. 10(b).

The experimental ratio $\Delta B_{2/5} / \Delta B_1 \approx 7.10 \pm 0.07$ is consistent with formation of an $r_{In} \approx 570$ nm, $f = 2/5$ island surrounded by an $r_{Out} \approx 685$ nm, $f = 1/3$ ring. Figure 3 shows the $B = 0$ 2D electron density $n(r)$ as a function of radius $r$ in the island, calculated following Ref. 23. Since electrostatic energy of electrons in the confining potential is large, $n(r)$ is expected to follow closely even in quantizing $B$. The opening of the quantum Hall gaps in the island leads to formation of the "compressible" and "incompressible" rings, with small density variations from the $B = 0$ profile[28]. The blue circle gives the radius of the outer edge ring, $r_{Out} \approx 685$ nm, obtained from the Aharonov-Bohm period data for $f_B = 1$ and 2, shown in Fig. 6. The magenta circle gives the radius of the inner edge ring, $r_{In} \approx 570$ nm for $f = 2/5$ from the data of Figs. 9 and 10(a). The ordinate of the magenta circle is obtained from the ratio of densities $n(r_{In})/n(r_{Out}) = (2/5)/(1/3) = 1.20$, taking the same $r_{Out}$ for $f = 1/3$ as above. The fact that the magenta circle is very close to the calculated $n(r)$ profile illustrates the reasonableness of the inner-outer ring edge assignment. The resulting width of the $f = 1/3$ ring, $r_{Out} - r_{In} \approx 115$ nm (15 magnetic lengths), is also reasonable for edge channel separation.[23,28]

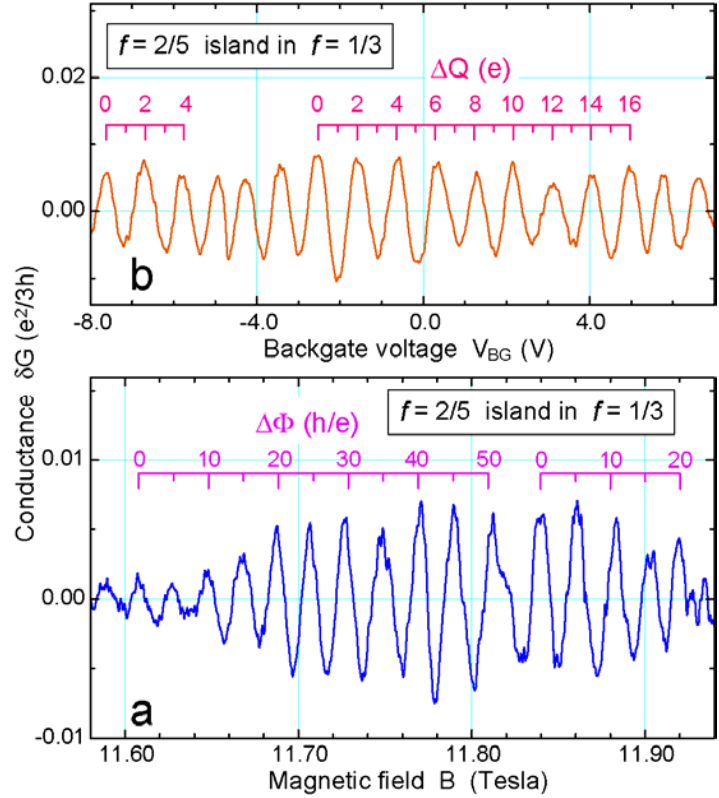

**Fig. 10** (Color online) Interference of the inner ring $e/3$ Laughlin quasiparticles circling an island of the $f = 2/5$ FQH fluid. (a) Magnetic flux through the island period of $\Delta\Phi = 5h/e$ corresponds to creation of ten $e/5$ quasiparticles in the island (one fundamental flux quantum $h/e$ induces two quasiparticles in the $f = 2/5$ FQH fluid). Such "superperiod" of $\Delta\Phi > h/e$ has never been reported before. (b) The backgate voltage period of $\Delta Q = 10(e/5) = 2e$ directly confirms that the $e/3$ quasiparticle consecutive orbits around the island are quantized by the condition requiring incremental addition of ten $e/5$ quasiparticles of the $f = 2/5$ fluid. These observations imply relative fractional statistics, when a charge $e/3$, statistics $\Theta_{1/3} = 2/3$ quasiparticle encircles one $e/5$, $\Theta_{2/5} = 2/5$ quasiparticle of the $f = 2/5$ fluid.

The ratio of the conductance oscillations periods is determined by the quantum Hall filling, independent of the edge ring area: $\Delta B / \Delta V_{BG} \propto N_\Phi / N_e = 1/f$, where $N_\Phi$ and $N_e$ are the number of flux quanta and electrons, respectively, in the area of the encircled path. The fact that the ratios fall on a straight line forced through zero confirms the island filling as $f = 2/5$, see Figure 11. Note that island filling assignments of either 1/3 or 3/7 (the neighbors of 2/5 in the FQH sequence) are ruled out by the data of Fig. 11. The island $f = 2/5$ assignment of the data is further supported by three additional considerations. (i) We have observed similar integer and fractional Aharonov-Bohm data in another interferometer sample,[29] with a larger lithographic radius $R \approx 1{,}300$ nm. The ratio $\Delta B_{2/5} / \Delta B_1 \approx 6.4$ is consistent, upon the same depletion potential analysis, with the island filling assignment of $f = 2/5$; the fact that the ratio is less than 7 rules out the 3/7 assignment. (ii) We do observe what we believe to be the oscillations when the center island filling is $f = 3/7$ at lower magnetic fields. The structure of these oscillations is not sufficiently simple to be reported and/or analyzed at this time. (iii) The island $f = 2/5$ surrounded by an $f = 1/3$ ring is the most simple possible configuration, and involves the two strongest (largest gap) FQH fluids. It would be odd to observe a more subtle and weaker effect, and (despite numerous measurements under varying conditions) not to observe the simpler and stronger one.

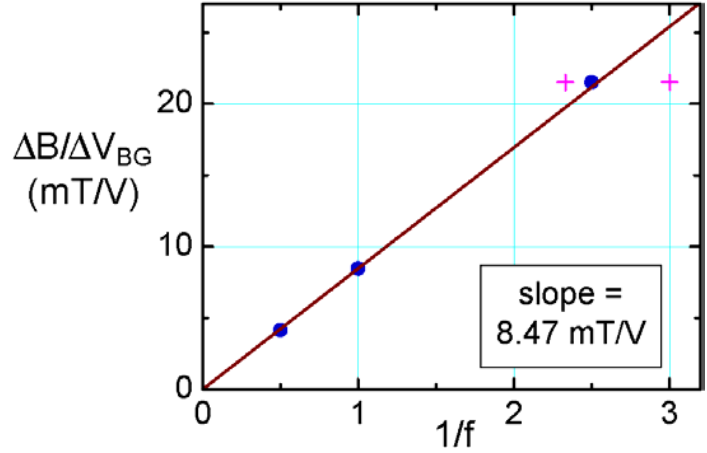

**Fig. 11** (Color online) The ratio of the conductance oscillations periods $\Delta B / \Delta V_{BG} \propto 1/f$, which is independent of the edge ring area, from the data shown in Figures 6, 7 and 10. The fact that the ratios fall on a straight line forced through (0,0) and the $f=1$ point confirms the island filling $f=2/5$. The magenta crosses give the neighboring $f=3/7$ and $f=1/3$, which clearly do not fit the data.

### V. Fractional statistics of quasiparticles

The striking feature of the conductance oscillations shown in Figure 10(a) is that the Aharonov-Bohm period is five fundamental flux quanta: $\Delta\Phi = 5h/e$! To the best of our knowledge, such "superperiod" of $\Delta\Phi > h/e$ has never been reported before. The gauge invariance argument of the Byers-Yang theorem[18] requiring $\Delta\Phi \leq h/e$ for the true Aharonov-Bohm geometry is not applicable in the present sample because the *interior* of the edge ring contains electrons.

Addition of magnetic flux $h/e$ through an area occupied by the $f=1/3$ FQH fluid creates a vortex in the many electron wave function, that is, a charge $e/3$ Laughlin quasihole.[4,8] Similarly, addition of flux $h/e$ to the $f=2/5$ FQH fluid creates two vortices in the many electron wave function, that is, two charge $e/5$ quasiholes.[30,10] These theoretical predictions have been verified at a microscopic level in the quantum antidot experiments.[6,13,14] Thus, addition of a flux $\Delta\Phi = 5h/e$ to the $f=2/5$ island creates ten $e/5$ quasiparticles with total charge $\Delta Q = 2e$. It should be noted that addition of flux into the island does not affect the total charge: excitation of quasiparticles out of a condensate leaves total electronic charge constant. For example, increasing magnetic field so that there is one more flux quantum through the island will excite two $e/5$ quasiholes, and, at the same time, reduce the electronic (negative) charge of the exact filling condensate by precisely $2e/5$, thus leaving the total island charge unaffected.

In contrast, a gate voltage does repel or attract 2D electrons. Here, magnetic field is fixed, so that the condensate charge is fixed too, and creation of quasiparticles does change the total charge of the island. The remote backgate produces relatively uniform electric field (compared to the field of the front gates). As demonstrated by the integer QH data of Fig. 7, a small variation of the backgate voltage can change the charge of the island, while leaving the island on the same QH plateau. In the fractional QH regime, the $\Delta Q = 2e$ charge periodicity, corresponding to creation of ten $e/5$ quasiparticles, is directly confirmed by the backgate data in Figure 10(b). In contrast, the charge periodicity observed in quantum antidots corresponds to addition of *one* quasiparticle only, both for the $f = 1/3$ and $f = 2/5$ cases. The principal difference between the present interferometer and the quantum antidots is that while in quantum antidots the FQH fluid surrounds electron vacuum,[17] in the present interferometer the 1/3 fluid surrounds an island of the 2/5 fluid.

If we neglect the symmetry properties of the FQH condensates, in the absence of a Coulomb blockade, there is no *a priori* requirement that the total charge of the FQH island be quantized in units of $e$, much less in units of $2e$. The island charge could change in increments of one quasiparticle charge, any small (less than $e$) charge imbalance supplied from the contacts. This is so because the island is a part of a larger electron system, the surrounding FQH fluid being connected by Ohmic contacts and wires to the rest of the world. As is well known, in an open system the chemical potential is fixed, not the number of particles. Ref. 20 considers an isolated (channel + island) system, where the total number of electrons is fixed. Their model has an additional constraint, that no quasiparticles are created, thus leading to formation of a charged ring at the 1/3 – 2/5 boundary and large Coulomb energy. For example, for the tenth fringe from the exact filling, the ring net charge would be $20e$, and the charging energy ~1,000 K for the ring radius of 600 nm. This energy should be compared to ~1 K needed to excite an $e/5$ quasiparticle.[31,32] Thus, inhibition of quasiparticle creation is not energetically possible, and does not describe the FQH ground state in the island away from the exact filling. Inhibition of quasiparticle creation is an essential basis of their derivation, thus their results are not directly applicable to this experiment.

Therefore, the quantization periods of $\Delta\Phi = 5h/e$ and $\Delta Q = 2e$ must be imposed by the symmetry properties of the two FQH fluids. The current used to measure conductance is transported by the quasiparticles of the outside 1/3 fluid; thus, we must construe that the conductance oscillations periodicity of ten $e/5$ quasiparticles results from the $2\pi$ periodicity of the Berry phase of a charge $q = e/3$ Laughlin quasiparticle encircling ten $e/5$ quasiparticles of the $f = 2/5$ fluid:

$$\Delta\gamma = \frac{q}{\hbar}\Delta\Phi + 2\pi\Theta\,\Delta N = (\frac{e}{3\hbar})(\frac{5h}{e}) + 2\pi\Theta_{2/5}^{1/3}(10) = 2\pi \ . \tag{6}$$

Solving Eq. (6) gives the relative statistics $\Theta_{2/5}^{1/3} = -1/15$.

The elementary texts usually define exchange statistics of *identical* particles. The notion of *relative* statistics of non-identical anyons should not be surprising, though, specifically for Laughlin quasiparticles, since all elementary charged excitations of various FQH fluids fundamentally are collective excitations of strongly interacting 2D electrons. Wilczek has considered "mutual" fractional statistics of quasiparticles in a two-layer FQH system with unequal fillings;[33] Su *et al*. have considered mutual exclusion statistics between quasielectrons and quasiholes of the same FQH condensate.[34] Ref. 20 derives statistics of quasiparticles of the channel FQH fluid only; their result given in Eq. (15) is equal (mod 2) to their Eq. (3), which is consistent with our Eq. (3) for $j=1$. We are not aware of a theoretical work containing an explicit evaluation of $\Theta_{2/5}^{1/3}$.

Thus, the microscopic mechanism leading to $\Delta\Phi = 5h/e$ and $\Theta_{2/5}^{1/3} = -1/15$ is not fully understood at present. It may be tempting to explore the fact that $f_C - f_B = 1/3 - 2/5 = -1/15$ happens to be equal to the experimental value of $\Theta_{2/5}^{1/3}$; likewise, $\Theta_{1/3} + \Theta_{2/5} - 1 = 2/3 + 2/5 - 1 = 1/15$. Such arithmetic is not transparent from the underlying physics. In particular, the charge of the elementary excitations of $f = 2/5$ is $e/5$, not $2e/5$. Indeed, Eqs. (3) and (5) require input of values of quasiparticle charge $q$ and quasiparticle degeneracy $p$, the two properties that characterize a particular simple FQH condensate; filling factor alone does not *a priory* provide all the necessary information. The arithmetic in Eq. (6) is different: assuming the experimental $\Delta\Phi = 5h/e$, then $\Theta_{2/5}^{1/3} = (q_{2/5} - q_{1/3})/p_{2/5} = (1/5 - 1/3)/2 = -1/15$, where $p_{2/5} = 2$ is the $e/5$ quasiparticle degeneracy, *cf.* Eq. (3). The principal experimental results of $\Delta Q = 2e$ and $\Delta\Phi = 5h/e$ remain unexplained. In principle, $\Delta\Phi$ can be evaluated theoretically either in numerical work, or by an analytical calculation similar to Ref. 8, using unprojected composite fermion wave functions.

It is quite evident that Eq. (6) can be satisfied neither by bosonic nor fermionic integer $\Theta_{2/5}^{1/3}$ statistics, therefore an exchange of charge between the island and the surrounding FQH fluid in increments of one quasiparticle charge, $\Delta Q = e/5$, is not possible even in the absence of Coulomb blockade. A naive argument that charge transfer between the island and the surrounding fluid may always proceed in increments of one electron charge, $\Delta Q = e$ does not take into account the statistical phase contribution and is not in accord with the experiment. It is also easy to see that no physically meaningful singular gauge transformation would restore a $\Delta \Phi = h/e$ flux periodicity in this system. The central experimental results obtained, that is, the oscillations periods of $\Delta \Phi = 5h/e$ and $\Delta Q = 2e$, are robust and do not involve any adjustable-parameter fitting to a theory. Thus we conclude that the experiment reported here provides a direct and unambiguous observation of fractional statistics of FQH quasiparticles.

## VI. Outlook: topological quantum computation

We have realized a novel Laughlin quasiparticle interferometer where the wave function of quasiparticles encircling a FQH fluid island acquires a fractional statistical phase. This experiment opens a new regime in the many-body physics of interacting particles confined to move in two dimensions. The fractional statistics quasiparticles, the anyons, are of interest not only in a fundamental science, but yet may find a practical application in quantum information processing. Environment-induced decoherence and the unavoidable spread of qubit parameters present the two most significant obstacles to practical implementation of scalable solid-state quantum logic circuits. Topological quantum computation with abelian and non-abelian anyons has been suggested as a way of implementing intrinsically fault-tolerant quantum computation.[35-37] Intertwining of anyons with non-trivial exchange statistics induces unitary transformations of the system wave function that depend only on the topological order of the underlying FQH condensate.[24] These transformations can be used to perform quantum logic, the topological nature of which is expected to make it more robust against environmental decoherence.

An experimentally feasible approach proposes adiabatic transport of FQH quasiparticles in systems of quantum antidots for implementation of the basic elements for anyonic quantum computation.[38] The basic qubit consists of a quantum antidot "molecule" occupied by one

"extra" quasiparticle.[39] In the experimentally realizable low temperature, low electromagnetic environment limit, modulation of front gates' potentials can be used to attract quasiparticles one by one to an antidot. Here, computation employs a fractional Berry phase created by an adiabatic transfer of one quasiparticle around another in systems of quantum antidots to perform quantum logic. The key general question yet to be answered in the future work is: to what extent the topological nature of the statistical phase of quasiparticles helps to alleviate the decoherence in quantum computation?

**Acknowledgements:** We thank D. V. Averin for discussions. This work was supported in part by US NSA and ARDA through US Army Research Office under Grant DAAD19-03-1-0126 and by the National Science Foundation under Grant DMR-0303705.

**References:**

1. J. M. Leinaas and J. Myeheim, Nuovo Cimento Soc. Ital. Fis. **37B**, 1 (1977).

2. F. Wilczek, Phys. Rev. Lett. **48**, 1144 (1982); Phys. Rev. Lett. **49**, 957 (1982).

3. Y. Aharonov and D. Bohm, Phys. Rev. **115**, 485 (1959).

4. R. B. Laughlin, Phys. Rev. Lett. **50**, 1395 (1983); Rev. Mod. Phys. **71**, 863 (1999).

5. D. C. Tsui, H. L. Stormer, and A. C. Gossard, Phys. Rev. Lett. **48**, 1559 (1982).

6. V. J. Goldman and B. Su, Science **267**, 1010 (1995); Physica E **1**, 15 (1997).

7. B. I. Halperin, Phys. Rev. Lett. **52**, 1583 (1984).

8. D. Arovas, J. R. Schrieffer, and F. Wilczek, Phys. Rev. Lett. **53**, 722 (1984). Ref. 8 defines clockwise as the positive direction of circulation, thus their statistics is $-\Theta$; e.g., their quasihole statistics $\Theta_{1/3} = 1/3$ is $-1/3 = 2/3 \mod 1$ in right-handed coordinates

9. M. V. Berry, Proc. Royal Soc. A **392**, 45 (1984).

10. W. P. Su, Phys. Rev. B **34**, 1031 (1986).

11. While Landau level filing factor $\nu \equiv hn/eB$ is a variable, the QH exact filling $f$, defined as inverse of the value of the *quantized* Hall resistance $R_{XY}$, in units of $h/e^2$ (that is, $f \equiv h/e^2 R_{XY}$), is a quantum number.

12. For the FQH hierarchy, see J. K. Jain and V. J. Goldman, Phys. Rev. B **45**, 1255 (1992).

13. V. J. Goldman, Surf. Science **362**, 1 (1996).

14. V. J. Goldman, I. Karakurt, J. Liu, and A. Zaslavsky, Phys. Rev. B **64**, 085319 (2001).

15. L. Saminadayar, D. C. Glattli, Y. Jin, and B. Etienne, Phys. Rev. Lett. **79**, 2526 (1997).

16. R. De-Picciotto *et al.*, Nature **389**, 162 (1997).

17. We consider here the model of a quantum antidot as a dense disc of quasiholes on top of (and surrounded by) a quantum Hall condensate. See A. H. MacDonald, Science **267**, 977 (1995); V. J. Goldman, J. Liu, and A. Zaslavsky, Phys. Rev. B **71**, 153303 (2005).

18. N. Byers and C. N. Yang, Phys. Rev. Lett. **7**, 46 (1961); C. N. Yang, Rev. Mod. Phys. **34**, 694 (1962).

19. S. A. Kivelson, Phys. Rev. Lett. **65**, 3369 (1990).

20. J. K. Jain, S. A. Kivelson, and D. J. Thouless, Phys. Rev. Lett. **71**, 3003 (1993).

21. C. de C. Chamon *et al.*, Phys. Rev. B **55**, 2331 (1997).

22. I. J. Maasilta and V. J. Goldman, Phys. Rev. B **55**, 4081 (1997).

23. B. Y. Gelfand and B. I. Halperin, Phys. Rev B **49**, 1862 (1994).

24. X. G. Wen, J. Mod. Phys. B **6**, 1711 (1992).

25. R. J. Haug, A. H. MacDonald, P. Streda, and K. von Klitzing, Phys. Rev. Lett. **61**, 2797 (1988); S. Washburn, A. B. Fowler, H. Schmid, and D. Kern, Phys. Rev. Lett. **61**, 2801 (1988).

26. L. P. Kouwenhoven *et al.*, Surf. Science **229**, 290 (1990).

27. I. Karakurt, V. J. Goldman, J. Liu, and A. Zaslavsky, Phys. Rev. Lett. **87**, 146801 (2001).

28. D. B. Chklovskii, B. I. Shklovskii, and L. I. Glazman, Phys. Rev B **46**, 4026 (1992).

29. Unfortunately, the second interferometer sample displaying similar data was destroyed by an electrostatic discharge before an extensive data set in the FQH regime could be acquired. The integer regime data is reported in F. E. Camino, W. Zhou, and V. J. Goldman, cond-mat/0503456.

30. F. D. M. Haldane, Phys. Rev. Lett. **51**, 605 (1983).

31. R. H. Morf, N. d'Ambrumenil, and S. Das Sarma, Phys. Rev. B **66**, 075408 (2002).

32. R. R. Du *et al.*, Phys. Rev. Lett. **70**, 2944 (1993).

33. F. Wilczek, Phys. Rev. Lett. **69**, 132 (1992).

34. W. P. Su, Y. S. Wu, and J. Yang, Phys. Rev. Lett. 77, 3423 (1996).

35. A. Y. Kitaev, Ann. Physics **303**, 2 (2003).

36. J. Preskill, Fault-tolerant quantum computation. In *Introduction to quantum computation and information* (Eds. H-K. Lo, S. Papesku, and T. Spiller) 213-269 (World Scientific, Singapore, 1998).

37. S. Lloyd, preprint at ⟨http://arXiv.org/quant-ph/0004010⟩ (2000).

38. D. V. Averin and V. J. Goldman, Solid State Commun. **121**, 25 (2002).

39. I. J. Maasilta and V. J. Goldman, Phys. Rev. Lett. **84**, 1776 (2000).